\documentstyle[12pt,epsf,epsfig]{article}
\def\be{\begin{equation}}
\def\ee{\end{equation}}
\def\ba{\begin{eqnarray}}
\def\ea{\end{eqnarray}}

\def\bdm{\begin{displaymath}}
\def\edm{\end{displaymath}}

\def\bq{\begin{quote}}
\def\eq{\end{quote}}

 at 10truept

\newcommand{\beq}{\begin{equation}}
\newcommand{\eeq}{\end{equation}}
\newcommand{\bea}{\begin{eqnarray}}
\newcommand{\eea}{\end{eqnarray}}
\newcommand{\beqa}{\begin{eqnarray}}
\newcommand{\eeqa}{\end{eqnarray}}

\newcommand{\lmk}{\left(}  \newcommand{\rmk}{\right)}
\newcommand{\lkk}{\left[}  \newcommand{\rkk}{\right]}

\newcommand{\calr}{{\cal R}}

\def\ltap{\ \raise.3ex\hbox{$<$\kern-.75em\lower1ex\hbox{$\sim$}}\ }
\def\gtap{\ \raise.3ex\hbox{$>$\kern-.75em\lower1ex\hbox{$\sim$}}\ }
\def\gl{\ \raise.5ex\hbox{$>$}\kern-.8em\lower.5ex\hbox{$<$}\ }
\def\roughly#1{\raise.3ex\hbox{$#1$\kern-.75em\lower1ex\hbox{$\sim$}}}

\begin{document}

\thispagestyle{empty}
\begin{flushright}
arXiv:0803.3809 [hep-ph]\\
RESCEU-04/08\\
August 2008
\end{flushright}
\vspace*{1.2cm}
\begin{center}
{\Large \bf Higgsflation at the GUT scale in a Higgsless Universe}\\

\vspace*{1.6cm} {\large Nemanja Kaloper$^{a,}$\footnote{\tt
kaloper@physics.ucdavis.edu}, Lorenzo Sorbo$^{b,}$\footnote{\tt
sorbo@physics.umass.edu} and Jun'ichi
Yokoyama$^{c,}$\footnote{\tt yokoyama@resceu.s.u-tokyo.ac.jp} }\\
\vspace{.5cm} {\em $^a$Department of Physics, University of
California, Davis, CA 95616, USA}\\
\vspace{.2cm} {\em $^{b}$ Department of Physics,
University
of Massachusetts, Amherst, MA 01003, USA}\\
\vspace{.2cm} {\em $^c$Research Center for the Early Universe (RESCEU),
 Graduate School of Science,\\
The University of Tokyo, Tokyo, 113-0033, Japan}\\

\vspace{1cm} ABSTRACT
\end{center}
We revisit inflation in induced gravity. Our focus is on models
where the low scale Planck mass is completely determined by the
breaking of the scaling symmetry in the field theory sector. The
Higgs-like field which breaks the symmetry with a GUT-scale {\it
vev} has non-minimal couplings to the curvature, induced by the
gravitational couplings of the other light fields in the theory,
so that its $vev$ controls the gravitational strength. This field
can drive inflation, and give a low energy universe in very good
agreement with the cosmological observations. The low energy
dynamics of the Standard Model cannot be unitarized by the
Higgsflaton, which decouples from the low energy theory, both
because it picks up a large mass and because its direct couplings
to the low energy modes are weakened. Instead, the short distance
behavior of the Standard Model may be regulated by the dynamics of
other light degrees of freedom, such as in Higgsless models.

\vfill \setcounter{page}{0} \setcounter{footnote}{0}
\newpage

The Standard Model of particle physics (SM) has been a singularly
successful framework for explaining the observed dynamics of
elementary particles. Its low energy canonical spectrum
contains\footnote{The discovery of the neutrino masses has already
taken us outside of the canonical SM, extending the spectrum by at
least as many as $6$ more fermions.} $90$ fermionic degrees of
freedom, $27$  vector bosons, with or without mass terms, and a
single scalar degree of freedom which has so far eluded all
attempts at detection. This evasive mode -- the Higgs field -- is
special in many respects. It is the only fundamental scalar in the
SM, having so far completely avoided direct detection. On the
other hand, the whole structure of the SM hinges on its existence,
because it is responsible for the unitarization of the electroweak
sector of the theory and the generation of vector boson and
fermion masses. Indeed, the Higgs spontaneously breaks the
electroweak gauge symmetry, setting the mass scale of the massive
SM fields\footnote{With the possible exception of neutrinos, which
by the see-saw mechanism may inherit masses from dynamics at much
higher scales.}. The residual Higgs fluctuations then regulate the
massive low energy electroweak sector and unitarize its scattering
amplitudes. However, not all is well in the SM tale. As is well
known, the Higgs mass is not radiatively stable, and its
phenomenological value of $\sim 100$~GeV, and consequently a flat
potential, begs the question about what may possibly keep it
there. This single missing SM degree of freedom is so vital to the
whole model that a spectacular machine such as the Large Hadron
Collider (LHC) has as one of its key tasks seeking for it, and for
the physics which makes it possible.

But, what if the LHC does not find a fundamental scalar? An
absence of the Higgs would push us into alternative explanations
of the observed low energy SM dynamics. Models without a light
Higgs may have to separate the origin of masses from the new
physics which unitarizes the low energy theory. Indeed, what would
(not) be directly observed at the LHC are only the fluctuations of
the Higgs field, and not its zero mode. Examples where the
unitarization of SM amplitudes is disentangled from the origin of
mass have already been considered, and among them recently the
Higgsless models in extra dimensions~\cite{higgsless} attracted
much attention. These models are realized as brane setups in
cutoff $AdS$ space \cite{RS}, with $AdS$ radius $L$, with fermions
localized on branes and gauge bosons propagating in the bulk. In
the effective $4D$ theory this yields towers of Kaluza-Klein gauge
bosons, whose masses and couplings are determined by the warping
of the bulk $AdS$ geometry. In the dual cutoff $AdS/CFT$, they are
duals of light $CFT$ states, with masses below the $CFT$ UV cutoff
$\mu \sim 1/L$, whose number is ${\cal N} \sim (M_4 L)^2$
\cite{nimalisa}. Curiously, this counting of light states agrees
with the recent ideas on a relation of the number of light states
below some UV cutoff and hierarchy between this cutoff and the
Planck scale \cite{gia}. In fact, such models are naturally
related to technicolor models, but now in the strong coupling
regime as defined by way of the $AdS/CFT$ duality, where the SM is
also unitarized without invoking fundamental scalars.

Thus if no fundamental scalar were observed at the LHC, we will
have had the (poor!) consolation of verifying experimentally the
theoretical prejudice against light fundamental scalars. Such an
outcome would speak loudly against the existence of {\em any}
light fundamental scalar, indicating that Nature may choose other
routes for realizing the low energy SM. We stress once again
here, that while this may be an extreme point of view, it is not
{\it yet} excluded by any experimental facts. SM with the Higgs is
the simplest means of describing the observed particle dynamics,
but the Higgs is still missing, and its existence has been
questioned. Thus it is of interest to consider other implications
of a missing Higgs, particularly as it relates to the paradigm of
naturalness. While speculative from the point of view of our
current expectations, they yet remain to be excluded.

Indeed, the absence of the Higgs would have other
implications if we take the view that its presence were to support
naturalness. Beyond the SM physics, scalars also play a key role
in cosmology, where they are prototypical inflatons. A scalar
field provides the simplest dynamics necessary to inflate the
universe, ensuring that its large smooth and nearly flat swaths
survive to the present epoch~\cite{inflation}. The scalars can be
inflatons if their potential is very flat in the units of the
natural cutoff, compared to which they are light. This is
qualitatively similar to the SM Higgs, although in practice the
requirement for potential flatness is quantitatively weaker.
Nonetheless, all inflaton models need some amount of fine tuning
to make sure the potential remains flat in spite of the couplings
to other matter, necessary for reheating.

In the event that LHC finally discovers the Higgs, it will be easy
to imagine that other light scalars with flat potentials appear in
Nature, regardless of why that may be so. One could be the
inflaton, at a scale well below the cutoff, and well separated
from UV physics, and ultimately gravity. On the other hand, if no
light scalars are seen, a logical consequence may be that light
scalars are hard to sustain. In such an instance, the scalars
would drift up to near the cutoff, which may be at the GUT scale
$\sim 2\times 10^{16}$~GeV, as hinted at already from
sub-TeV-scale physics explored so far, by the proton stability,
the see-saw explanation neutrino masses, and the clues from gauge
coupling unification\footnote{Some features of dynamics in
Higgsless models as pertaining to these scales, and specifically
issues of relevance for unification have been addressed in
\cite{HGuts}.}. So if the scalars can't be stabilized near the TeV
scale, radiative stability may be attained if the scalar masses
are pushed high, to near the GUT scale, where even the mode
responsible for inducing SM masses may end up. In such a scenario,
inflaton would be no exception. However unlikely this option may
now seem, the conspicuous absence of the Higgs from the observed
bestiary of elementary particles found to date points to the fact
this is not yet {\em impossible}. Indeed, the Higgsless models of
various kinds already account for this in the SM sector. In this
note, we shall outline how to allow for inflation in such a
Universe, basing it on a Higgs-like field, which spontaneously
breaks the scale symmetry at the GUT scale, and gets a flat enough
effective potential, as it induces the (small) hierarchy between
the GUT scale and the Planck scale.

In the early days of inflationary model building, the possibility
of driving inflation by the SM Higgs has been tried, but without
immediate success~\cite{newinflation}. With minimal couplings to
gravity, the SM constraints force the scalar self-couplings to be
too large to yield satisfactory inflationary density perturbations
(see~\cite{keith} for a review). These problems can be ameliorated
if non-minimal couplings to gravity are allowed. In particular, in
the induced gravity framework~\cite{sainduced,induced} one can get
the right density perturbations even if the scalar self-couplings
are much larger than in the minimal coupling case \cite{indinfl}.
Recently it has been noted \cite{shaposhnikov} that if the scalar
has direct coupling to the curvature, $\sim \xi |\phi|^2R$, and
there is also the standard Einstein-Hilbert term in the theory,
$\sim M_{Pl}^2 R$, then  the scalar could both drive a low scale
inflation, yielding the right density contrast, and serve as the
Higgs after inflation. For this claim, it is crucial that the
gravitational sector contains the Einstein-Hilbert
term\footnote{More aspects of this scenario were considered in
\cite{parkyama}.}. If it weren't so, the COBE normalization and
the phenomenologically required Higgs $vev$, $ \langle H \rangle
\sim 246$~GeV,  would force the value of the Planck scale to be at
the $\sim 10$~TeV scale.

Our route here is very different. We imagine that the underlying
theory is conformal in the UV, including the gravitational sector.
This means that the bare gravitational Einstein-Hilbert term is
absent from the action, which instead contains higher
derivative terms, starting with the curvature squared invariants,
\begin{equation}
{\cal S} =\int d^4 x \sqrt{g} \lkk A \, {\cal GB} + B
C_{\mu\nu\lambda\sigma}^2 + C R^2 -\frac12({\nabla}\phi)^2-
\frac{\lambda}{4}\bigl(\phi^2 - v^2\bigr)^2 - {\cal L}_{\tt
matter}(g^{\mu\nu}, \phi, \psi) + \ldots \rkk \, ,
\label{actionsqr}
\end{equation}
where ${\cal GB} = R_{\mu\nu\lambda\sigma}^2 - 4 R_{\mu\nu}^2 +
R^2$ is the Gauss-Bonnet combination, $C_{\mu\nu\lambda\sigma}$ is
the Weyl tensor, and $A, B, C$ are some dimensionless constants.
This theory is in fact renormalizable, as shown some time ago in
\cite{stelle} and, later, in works on induced gravity
\cite{induced} and relation between Newton's constant and scale
symmetry breaking \cite{fubini}. On the other hand, suppose there
is a somewhat large number of degrees of freedom in the matter
sector, $\sim {\cal O}(10^4)$, including those which will become
the low energy SM. If there is a gauge group in the theory which
confines at some scale, dimensional transmutation will yield an IR
cutoff, which will be fed back to the scalars. There may also be
explicit symmetry breaking terms in the scalar sector, with the
scalars which are not protected from radiative corrections from
the strong gauge group.

Either way, the matter sector quantum field theory will be
characterized by a dimensional cutoff. Then, the quantum one-loop
effects will generate contributions to the action of the form
$\sim \Lambda^2 R$ \cite{sainduced,induced}. In general,
these corrections will depend on the cutoff itself, as well as the
value of field {\it vevs} around which the corrections are
calculated. We will {\it assume} that the field independent
contributions to $\Lambda$ can be neglected. This could be
justified as follows. The quantum contributions to $\sim R$ term
will come as $(\Lambda^2 + c \phi^2) R$ from every degree of
freedom which couples to gravity. If these degrees of freedom are
all weakly coupled, one would expect that the bare cutoff terms
may dominate. On the other hand, if some are in strong
coupling, the strong coupling effects may conspire between
different orders in the loop expansion and retain the appearance
of conformality, such that the dimensional transmutation which
they trigger may occur at a scale well below the strong coupling
scale \cite{marstrass}. Thus this scale could be smaller than the
one directly sampled by the Higgsflaton symmetry breaking\footnote{At
least in the weak coupling this may occur, as we know from the
example of QCD, where low energy quark masses are mainly
attributed to the electroweak symmetry breaking.}. Then the
leading order contributions to $\Lambda$ may come from the IR
masses of the fields residing in the geometry, yielding by linear
superposition $\Lambda^2 \sim \sum_k m_k^2$. If these masses
are generated directly by a symmetry breaking induced by a
Higgs-like field (Higgs for short from now on), $m_k(\phi) \sim g
\phi$, this would yield $\Lambda^2 \simeq {\cal N} g^2 \phi^2$,
which can dominate over the hard cutoff contributions. Here for
simplicity we assume that all the Yukawa couplings are
approximately the same. The number ${\cal N}$ counts the fields in
the theory which are Higgsed by $\phi$, and so this yields $\xi
\sim {\cal N} g^2$. Again, this is consistent with the recent
ideas about the large number of light fields inducing the
hierarchy between the mass scale where they reside and the Planck
scale \cite{gia}, although it would be a much more conservative
quantitative implementation of such a framework.  Note, that
the crucial aspect of this idea is that the conformal symmetry
breaking which induces the Einstein-Hilbert term is {\it soft}, in
that the hard cutoff contributions must be subleading, which
typically may not occur in weak coupling \cite{marstrass}.

Of course, the scalar which breaks the symmetry cannot be the
usual Higgs~\cite{dehnen,shaposhnikov}, since its mass will be too
large, as would be natural by the low energy accounting of
radiative corrections. This scalar will have its mass and $vev$
set by the scale where the conformal symmetry breaks down. To
reflect this, we will dub it the ``Higgsflaton", and take the
symmetry breaking scale to be the GUT scale\footnote{The proximity
of the GUT scale to the Planck scale makes the presence of
fundamental scalars near the GUT scale appear more plausible,
since at those scales one may get away without mechanisms that
protect their masses from radiative corrections.}. However the
crucial property that allows the Higgsflaton to drive inflation,
and therefore get a somewhat flatter potential, is its coupling to
the Ricci scalar. The key reason is that the parameter $\xi$, of
the order of $10^4$, needed to induce the hierarchy between the
GUT scale and the Planck scale, also {\it see-saws} the scalar
mass from the GUT scale down to $m_\varphi \sim v/\sqrt{\xi}$,
flattening the scalar potential just enough. Moreover, this number
precisely reproduces the COBE normalization\footnote{Possible
connections between the GUT scale and primordial density
perturbations were noted in \cite{tomb}, albeit realizations were
different.} of the scalar density perturbations in this model.
Given the argument for how the Einstein-Hilbert term comes about,
the value of $\xi$ can be obtained by positing that the
Higgsflaton gives mass to about $10^4$ degrees of freedom, with
Yukawa couplings $g \sim 1/3$, which therefore live at the
GUT scale, and whose loops induce the Einstein-Hilbert term. In
this case, the low energy Standard Model is unitarized by some
other degrees of freedom, e.g. as in the Higgsless models
\cite{higgsless}.  Note that in this scenario -- as in the
Higgsless model -- we are not addressing the origin of the
electroweak scale, which should be attributed to some other strong
dynamics that does not necessarily involve scalar modes. At least
the SM fields, being outnumbered by the other degrees of freedom
in the theory, and much lighter than most, will not contribute
significantly to the generation of the Einstein-Hilbert term,
which would be largely insensitive to their presence.

Let us now outline the cosmological scenario. In light of
the discussion above, the low energy theory, below the scale
symmetry breaking, is given by the effective $4D$ action
\begin{equation}
{\cal S} =\int d^4 x \sqrt{g} \lkk\frac12\xi\phi^2 R
-\frac12({\nabla}\phi)^2- \frac{\lambda}{4}\bigl(\phi^2 -
v^2\bigr)^2 - {\cal L}_{\tt matter}(g^{\mu\nu}, \phi, \psi) + \ldots
\rkk \, , \label{action}
\end{equation}
where $ {\cal L}_{\tt matter}$ includes the Standard Model and
additional matter fields which unitarize it at the $\sim TeV$
scale, collectively denoted by $\psi$, and $\phi$ is the
Higgsflaton scalar field modulus, with a non-minimal coupling to
curvature $\xi \phi^2 R$. The ellipsis stand for additional terms
which we assume to be mostly negligible. The Higgsflaton phase is
in $ {\cal L}_{\tt matter}$ as a longitudinal component of a gauge
boson, so that $ {\cal L}_{\tt matter}$ is written in a unitary
gauge. With the assumptions above, the parameterization of its
leading order low energy dynamics by  (\ref{action}) is accurate
in the limit of weak gravity. On the other hand, although in the
regime of background field values $\phi \sim 0$ the field theory
in (\ref{action}) is perturbative, gravity as encoded by
(\ref{action}) becomes strong. So sufficiently close to the origin
in field space the theory cannot be described by (\ref{action}).
However, in this regime the scale symmetry is restored, and the
gravitational theory reverts back to the curvature squared action,
with a negligible Einstein-Hilbert correction.

At any rate, at low energies for large values of $\phi$ which
break the symmetry, gravity will be weak when $\xi \gg 1$. In this
limit, we can use the field equations derived from (\ref{action})
to describe the background geometry. At a minimum, $\phi = \pm v$,
if we integrate out the scalar the theory reduces to ${\cal
S}_{\tt eff} =\int d^4 x \sqrt{g} \left(\frac12\xi v^2  R - {\cal
L}^{\tt eff}_{\tt matter}(g^{\mu\nu}, \psi) + \ldots \right)$,
which shows that the effective low energy Planck scale around the
scalar vacuum is
\be
M_{Pl}^2 = \xi v^2 \, .
\label{plmass}
\ee
To see the scalar dynamics we can go to the unitary gauge where
all fields are canonically normalized. Taking the conformal
transformation and scalar field redefinition
\cite{indinfl,FM,kko},
\be \hat g_{\mu\nu} = \lmk \frac{\phi}{v} \rmk^2 \, g_{\mu\nu}
\, , ~~~~~~~~~ \varphi = M_{Pl} \, \sqrt{6+ \frac{1}{\xi}} \,
\ln\lmk\frac{\phi}{\phi_0}\rmk \, , \label{redefs} \ee
where $\phi_0$ is an arbitrarily chosen normalization, yields the
Einstein frame action
\ba && {\cal S} = \int d^4  x \sqrt{\hat g}
\left\{ \frac{M_{Pl}^2}{2} \hat R -{1\over2}(\hat \nabla\varphi)^2 -
\hat V(\varphi)\right.  \nonumber \\
&& ~~~~~~~~~~~~~~~~  -\left.  \lmk\frac{v}{\phi_0}\rmk^4 \,
e^{-4\frac{\varphi}{M_{Pl} \sqrt{6+1/\xi}}} {\cal L}_{\tt
matter}\lmk (\phi_0/v)^2 e^{2\frac{\varphi}{M_{Pl}
\sqrt{6+1/\xi}}} \, \hat g^{\mu\nu} , \varphi, \psi \rmk + \ldots
\right\} \, . \label{actefr} \ea
The new effective potential is, using Eq. (\ref{plmass}),
\be
\hat V(\varphi)= \frac{\lambda}{4}  \frac{(\phi^2-v^2)^2}{(\phi/v)^4} =
\frac{\lambda M_{Pl}^4}{4 \, \xi^2} \lkk 1- \lmk\frac{v}{\phi_0}\rmk^2
\, e^{-2\frac{\varphi}{M_{Pl} \sqrt{6+1/\xi}}}\rkk^2 \, .
\label{efrpot}
\ee
The minima $\phi = \pm v$ clearly correspond to $\varphi =  M_{Pl}
\, \sqrt{6+ \frac{1}{\xi}} \, \ln(\frac{v}{\phi_0})$. Around the
minimum, the curvature of the effective potential (\ref{efrpot})
yields the scalar mass
\be m^2_\varphi = \partial^2_\varphi \hat V = \frac{2\lambda \,
M_{Pl}^2}{\xi^2 (6 + 1/\xi)} \, , \label{scalmass} \ee
from which and (\ref{plmass}) it follows that
\be
m^2_\varphi \simeq \frac{\lambda v^2}{3\xi} \, ,
\label{mass}
\ee
in the limit  when $\xi \gg 1$. Obviously, in the limit $\xi \sim
1$, $m_\varphi \sim M_{Pl}$ and so this case is less interesting.
This is precisely the see-saw effect in the scalar sector, which
we alluded to in the introductory discussion. Indeed, that this is
akin to see-saw can be seen by eliminating $\xi$ from Eq.
(\ref{mass}) by using Eq. (\ref{plmass}), which yields
\be
m_\varphi^2 \simeq \frac{\lambda v^4}{3 M_{Pl}^2} \, ,
\label{seesaw}
\ee
precisely a see-saw mass formula. In fact, the dynamics
responsible for flattening the potential is conceptually similar
to scalar `seizing' of \cite{savscott}, except that the large
wavefunction renormalization involves the graviton as well as the
scalar field.

We note that the `strong gravity regime' $\phi \sim 0$ in the
Einstein frame variables corresponds to the limit $\varphi
\rightarrow - \infty$, where the potential (\ref{efrpot}), and
also all mass scales in the matter sector in (\ref{actefr})
diverge. This of course is simply the restatement of the fact that
the ratio of any mass scale $\mu$ and the effective Planck mass
$M_{Pl} = \sqrt{\xi} \phi$ diverges when $\phi \rightarrow 0$.
This manifestly excludes the limit $\varphi \rightarrow - \infty$
from the low energy action (\ref{actefr}), because in this case
one must restore the quadratic curvature terms which were ignored
in writing the effective action (\ref{action}).

For the potential (\ref{efrpot}), clearly inflation occurs when
$|\phi| > v$. In this limit gravity is weak, and furthermore the
potential behaves like a cosmological constant. This can be
readily seen from (\ref{efrpot}), since when $|\phi| > v$, $\hat V
\rightarrow \frac{\lambda v^4}{4} = \frac{\lambda
M_{Pl}^4}{4\xi^2}$. Thus, since the potential asymptotes a
constant when $\varphi \rightarrow \infty$, which smoothly goes to
the minimum $|\phi| = v$, sufficient inflation followed by a
graceful exit will occur when $\varphi$ is initially large. Note
however that by the formula (\ref{seesaw}), the mass of the
Higgsflaton at the minimum is comparable to the Hubble scale
during inflation, so the slow roll may extend even as the field
approaches the minimum. In the original variables, the initial
value of the field $\phi$ need not exceed $M_{Pl}$ when $\xi \gg
1$. This is qualitatively similar to assisted inflation
\cite{assisted}, where the expectation value of the inflaton
during inflation also need not be transplanckian. For more
complicated potentials, which may even involve bigger powers of
$\phi$, however, the effective potential (\ref{efrpot}) will still
typically have a maximum, and decay back to zero for very large
values of $\varphi$. In such cases, it is still possible to have
inflation if the initial value of $\varphi$ will be near the
maximum, which is expected to occur somewhere due to the random
distribution of initial values \cite{ekoy}.

Taking the background to be a spatially flat
Friedmann-Robertson-Walker spacetime, we can use the slow roll
equations to describe the geometry at large scales. This yields
\be \label{Feq1} {H}^2 \cong \frac{\lambda M^2_{Pl}}{12 \xi^2}
\lkk 1-\lmk\frac{v}{\phi}\rmk^2\rkk^2 \, , ~~~~~~~~~~ \dot\varphi
\cong - \frac{2}{\sqrt{3} \sqrt{6+1/\xi}}
\frac{\sqrt{\lambda}}{\xi} \lmk\frac{v}{\phi}\rmk^2 M_{Pl}^2 \, .
\ee
Using these equations (\ref{Feq1}), and recalling that curvature
perturbations are independent of the conformal frame in which they
are calculated \cite{kko,MS}, it is straightforward to determine
the amplitudes of scalar and tensor perturbations generate during
inflation. Their powers are, respectively,
\ba
\Delta_{\calr}^2&=&\lmk\frac{H^2}{2\pi\dot\varphi}\rmk^2
\cong \frac{\lambda}{128\pi^2\xi^2}
\lmk\frac{\phi}{v}\rmk^4\lkk 1-\lmk\frac{v}{\phi}\rmk^2\rkk^4 \, ,
\label{curpert}\\
\Delta_{h}^2&=& 8\lmk\frac{H}{2\pi M_{Pl}}\rmk^2 \cong
\frac{\lambda}{6\pi^2\xi^2}\lkk 1-\lmk\frac{v}{\phi}\rmk^2\rkk^2
\, , \label{tenpert} \ea
where we are taking the limit $\xi \gg 1$. Now, to determine the
scale at which (\ref{curpert}), (\ref{tenpert}) need to match to
the observed anisotropies, we need to relate the field values to
the inflationary clock readings, conveniently given by the number
of efolds before the end of inflation. Inflation will end when the
field rolls near the minimum $|\phi| \simeq v$. However to get a
precise location of the end of inflation, we can use the slow-roll
parameters in the Einstein frame, which for the potential
(\ref{efrpot}) are
\begin{eqnarray}
\epsilon &=& \frac{M_{Pl}^2}{2}\lmk\frac{\partial_\varphi \hat V}{
\hat V}\rmk^2 \simeq \frac{4}{3}\lkk
1-\lmk\frac{v}{\phi}\rmk^2\rkk^{-2}
\lmk\frac{v}{\phi}\rmk^4 \, , \label{slowr} \\
 \eta &=& M_{Pl}^2\frac{\partial^2_\varphi \hat V}{\hat V} \simeq
 \frac{4}{3} \lkk 1-\lmk\frac{v}{\phi}\rmk^2\rkk^{-2}
\lkk 2\lmk\frac{v}{\phi}\rmk^2-1\rkk
\lmk\frac{v}{\phi}\rmk^2 \, .\label{slowrr}
 \end{eqnarray}
Inflation will end when either $\epsilon$ or $\eta$ become ${\cal
O}(1)$. From (\ref{slowr}), this will occur at $|\phi|\equiv
\phi_* \simeq 1.47v$. This means that between some value
$\phi>\phi_\ast$ and this terminal value $\phi_\ast$, the universe
will undergo $N$ efolds of inflation, where $N$ is related to
$\phi$ according to
\be N= \int^\phi_{\phi_*} \frac{H}{\dot \varphi}
\frac{d\varphi}{d\phi} d\phi
 \cong
\frac{3}{4} \lmk\frac{\phi}{v}\rmk^2 -
\frac{3}{4} \lmk\frac{\phi_\ast}{v}\rmk^2
-\frac{3}{2}\ln\lmk\frac{\phi}{\phi_\ast}\rmk
\simeq \frac{3}{4} \lmk\frac{\phi}{v}\rmk^2 - 1-
\frac{3}{2}\ln\lmk\frac{\phi}{\phi_\ast}\rmk
 \, ,
\label{efolds}
\ee
where we have used (\ref{redefs}), (\ref{slowr}), and
(\ref{slowrr}). Since the pivot scale where CMB observations are
matched to the theory is $N_p = 55$, this implies that the
formulas for amplitude of perturbations (\ref{curpert}) and
(\ref{tenpert}) read
\be
 \Delta_{\calr}^2
\simeq \frac{\lambda}{72\pi^2\xi^2}
(N + 4.3)^2 \, ,  ~~~~~~~~~~~~~
 \Delta_{h}^2
\simeq \frac{\lambda}{6\pi^2\xi^2} \, ,
\label{expperts}
\ee
for $N\simeq 55$.  They yield $\Delta^2_{\cal R} \simeq 4.9 \times
\frac{\lambda}{\xi^2}$ and the tensor-to-scalar ratio
$r={\Delta_h^2}/{\Delta^2_{\cal R}} \simeq 0.003$.  This is within
the reach of future observational confirmation by planned
experiments of B-mode polarization observation of CMB such as
B-Pol. Matching the curvature perturbations to the observed value
of $\Delta_{\calr}^2=2\times 10^{-9}$ gives
\be
\frac{\sqrt{\lambda}}{\xi} \simeq 2.0 \times 10^{-5} \, .
\label{ratiocoupls}
\ee
We can also easily calculate the spectral index of the scalar
perturbations. The standard formula $n_s = 1 +  \frac{d \ln
\Delta^2_{\cal R}}{d \ln k}$ gives
\be
n_s \cong 1 - \frac{2}{N + 4.3} \, ,
\label{spectind}
\ee
which translates numerically to $n_s = 0.97$, in excellent
agreement with the CMB data. Aspects of the CMB constraints on the
perturbations in the model based on (\ref{action}) were also
considered in \cite{futkom}.

What of particle physics scales in this theory? As it manifest
from Eq. (\ref{ratiocoupls}), inflationary dynamics constrains the
ratio of the coupling constants $\lambda$ and $\xi$. To break this
degeneracy we can take the coupling $\lambda$ to be perturbative,
but not tiny, in order to relax the usual severe tunings in the
field theory sector of the inflaton \cite{indinfl}. So, suppose
that $\lambda \sim 10^{-2}$. In this case, Eq. (\ref{ratiocoupls})
implies $\xi \sim 5000$, and so by Eq. (\ref{plmass}) we find
\be
v = \frac{M_{Pl}}{\sqrt{\xi}} \sim 3\times 10^{16} \, \rm GeV \, ,
\label{vev}
\ee
i.e. we find $v \simeq M_{GUT}$, exactly as we asserted in the
introductory discussion. In turn the Higgsflaton mass
(\ref{scalmass}) in the vacuum $|\phi| =  v$ is $m_\varphi =
\sqrt{\lambda} \ v/\sqrt{3\xi} \sim  3\times 10^{13} \,$ GeV, by
Eq. (\ref{mass}), which thanks to the see-saw induced by the large
parameter $\xi$ is  significantly below the symmetry breaking
scale $v$.

As the field rolls down the slope of (\ref{efrpot}) towards the
minimum, it passes through an inflection point and the local
curvature of the potential, negative up on the plateau, will
increase slowly, eventually ending inflation. After falling down
the precipice to the potential well around the minimum, the field
oscillates around it on a time scale of the order of
$m_\varphi^{-1}$, reheating the universe. The details of reheating
depend on the couplings of the Higgsflaton to matter. The simplest
case is when in the original, Jordan frame, $\phi$ couples the SM
fermions only via Yukawa couplings. In this case, the classical
scaling symmetry allow us to completely decouple the
canonical Higgsflaton field $\varphi$ from matter. To see this
consider the transformation of the Jordan frame Yukawa term
$\sqrt{{g}}\,\phi\bar{\psi}\psi$ under conformal transformation.
The fermions will scale according to $\psi =
\left(\phi/v\right)^{3/2}\,\Psi$, so that $\Psi$ is canonically
normalized, which turns Yukawa couplings into simple mass terms
$\sqrt{\hat{g}}\,v\bar{\Psi}\Psi$ \cite{watako}. Without other
direct couplings of $\varphi$ to matter, reheating may occur in
two stages. In the first stage, $\varphi$ oscillates about the
bottom of its potential and its self interactions rapidly lead to
resonant amplification of the nonzero modes of $\varphi$, which
rescatter on the surviving part of the condensate, eventually
disrupting it \cite{Kofman:1997yn,tachpreh}, and ensuring that the
Universe is filled almost exclusively by quanta of $\varphi$ with
a typical momentum of ${\cal {O}}\left(m_\varphi\right)$.
Subsequently, the quanta of $\varphi$ will scatter against each
other, in processes like $\varphi\varphi\rightarrow \Psi\Psi$
mediated by gravitons, and produce the SM matter. A typical
timescale for this process the gravitational scattering scale
$\tau_{\tt {gs}}\simeq M_{Pl}^4/m_\varphi^5$, which with the mass
scale $m_\varphi\sim 3 \times 10^{13}$~GeV yields a reheating
temperature $T_{RH}\sim g_*^{-1/4}\,(M_{Pl}/\tau_{\tt
{gs}})^{1/2}\sim$~TeV.

In reality, however, the reheating will be more efficient, because
there will be additional couplings. To start with, one loop
corrections will the exact cancellation between the rescaling
factors in Yukawa terms, yielding a leftover field-dependent mass
$m_\Psi \sim v (\phi/v)^{d}$, where $d \sim {\cal O}(1)
\frac{g^2}{4\pi^2}$ arises from the anomalous dimension of the
fermions and the running of the coupling $g$. Thus the coupling
will in reality become $m_\Psi \sim v [1+ {\cal O}(1)
\frac{g^2}{4\pi^2} \ln(\frac{\phi}{v})]$, or after introducing the
canonically normalized field $\varphi$ from Eq. (\ref{redefs}) and
using $\xi \gg 1$,
\be
m_\Psi \sim v \, \Bigl(1+  {\cal O}(1) \frac{g^2}{4\pi^2} \frac{\varphi}{M_{Pl}} \Bigr)  \, .
\label{fermionmass}
\ee
This means that there will be Planck-suppressed couplings between
$\varphi$ and the fermions, and so the fermions will be produced
directly by the Higgsflaton tachyonic preheating, and additional
preheating stages as the field oscillates around the minimum
\cite{Kofman:1997yn,tachpreh}.

Moreover, if there are fields with explicit mass terms in the
theory, there will be  mass-term induced  direct couplings of
$\varphi$ to them, which are Planck-suppressed, but may still be
sufficiently large. This is most simply illustrated with an
example of a scalar field ${\chi}$ defined by a Jordan-frame
Lagrangian ${\cal {L}}_\chi= \sqrt{g}\,\left[ -\left(\nabla
\chi\right)^2/2 - U\left(\chi\right)\right]$. Upon changing to the
Einstein frame metric variable, we find the leading order
effective Lagrangian for $\chi$,
\be {\cal{L}}_{\tt eff}(\hat
\chi)=\sqrt{\hat{g}}\,\left[-\frac12\left(\hat \nabla
\chi\right)^2
\left(\frac{v}{\phi_0}\right)^2\,e^{-2\frac{\varphi}{M_{Pl}\sqrt{6+1/\xi}}}\,
-
\left(\frac{v}{\phi_0}\right)^4\,e^{-4\frac{\varphi}{M_{Pl}\sqrt{6+1/\xi}}}\,U\left(\chi\right)
+ \ldots \right] \, . \label{efflagrdec} \ee
If we expand this action in a series in $\varphi$ around the
minimum, where $\phi = \pm v$, and focus on the lowest order
terms, we can see that the trilinear Lagrangian describing lowest
order interactions is formed from keeping the kinetic term and the
mass term for $\chi$ and the linear term in $\varphi$. This yields
\be
{\cal{L}}_{\tt I} =\sqrt{\hat{g}}\,\frac{\varphi}{M_{Pl}\sqrt{6+1/\xi}} \,
\left[(\hat \nabla \chi)^2 + 2 m^2_\chi \chi^2  + \ldots
\right] \, .
\label{efflagrint}
\ee
Clearly, this trilinear term will yield the dominant channel for
$\varphi$ decay. The decay rate can now be calculated
straightforwardly. Since one is interested at the decay of
wavepackets much smaller than the Hubble length, one can ignore
the expansion of the universe and go to the locally Lorentzian
frame, by replacing the metric in (\ref{efflagrint}) by the
Minkowski metric. Then since the leading order process is $\varphi
\rightarrow 2 \chi$, one can go to the momentum picture and
evaluate (\ref{efflagrint}) on shell. That yields ${\cal{L}}_{\tt
I} = \frac{2 m^2_\chi - p_1 \cdot
p_2}{M_{Pl}\sqrt{6+1/\xi}}\varphi \chi^2 $, where $p_k$ are the
4-momenta of the decay products, and in the CM frame of $\varphi$
it reduces to, by recalling our metric signature to be $-+++$ and
using energy momentum conservation that yields $- p_1 \cdot p_2 =
\frac{m^2_\varphi}{2} - m^2_\chi$, an effective trilinear
interaction
\be {\cal{L}}_{\tt I} = \frac{m_\chi^2 +
m_\varphi^2/2}{M_{Pl}\sqrt{6+1/\xi}} \, \varphi  \chi^2 \, ,
\label{efftril} \ee
which is just the standard scalar Yukawa term with the coupling
constant $g = \frac{m_\chi^2 +
m_\varphi^2/2}{M_{Pl}\sqrt{6+1/\xi}}$. Therefore the decay rate
$\varphi \rightarrow 2\chi$ is\footnote{We are correcting here a
typo involving a sign in the formula for $g \sim m_\chi^2 +
m^2_{\varphi}/2$ in \cite{kko}.}
\be \Gamma_{\varphi \rightarrow 2\chi} = \frac{g^2}{8\pi
m_\varphi} \sqrt{1 - 4 \frac{m_\chi^2}{m_\varphi^2}} =
\frac{(m_\chi^2 + m_\varphi^2/2)^2}{8\pi (6+1/\xi) M_{Pl}^2
m_\varphi} \sqrt{1 - 4 \frac{m_\chi^2}{m_\varphi^2}} \, .
\label{decrat} \ee
Thus the gravitational decay time\footnote{The suppression of the
decay rate by $M_{Pl}$ arises since, by Eq. (\ref{redefs}), the
canonically normalized field $\varphi$ is multiplied by
$M_{Pl}^{-1}$. This agrees with the revised version
of~\cite{watako}.} when  $m_{\hat \chi} \simeq m_\varphi$ and $\xi
> 1$ is $\tau_{\tt {gd}}\sim
M_{Pl}^2\,m_\varphi/m_\chi^4$ \cite{kko}. These extra channels
will enhance reheating, and raise the reheating temperature: e.g.
if $m_{\hat \chi} \simeq m_\varphi\simeq 3\times 10^{13}$~GeV,
then $T_{RH}\sim 10^8$~GeV. The reheating temperature of this
range can directly be measured by observation of future
space-based laser interferometers \cite{nakayama}. Moreover, in
the presence of additional particles lighter than $\varphi$ new
channels will appear, enhancing $\Gamma \rightarrow \Gamma_{\tt
total} = \sum_k \Gamma_k$. In any case, the Higgsflaton will
settle down into the minimum rather efficiently. This is in fact
good, because if any energy in it survived, it could overclose the
universe. At any rate, this shows that the decay of the
Higgsflaton would be efficient, and will convert the vacuum-like
energy density of the Higgsflaton sector into normal particles.
The precise details would of course depend on the exact structure
of the physics which completes the Standard Model.

In lieu of a conclusion, let us state here that much of the
dynamics presented here will remain a possibility for inflation
even if LHC discovers the Higgs. In that case, however, many more
theories involving light scalars may be plausible, and when
identifying which may be the {\it raison d'etre} behind the
inflaton, one may fall victim to a `tyranny of small decisions'.
The absence of the Higgs could, at least in this sense, help
reduce the number of options for what lurks beyond the Standard
Model, and point to a high scale inflation, that could be tested
in the future searches for the primordial gravitational waves.

%%%%%%%%%%%%%%%%%%%%%%%%%%%%%%%%%%%%%%%%%%%%%%%%%%
\vskip0.5cm
%%%%%%%%%%%%%%%%%%%%%%%%%%%%%%%%%%%%%%%%%%%%%%%%%%%

{\bf \noindent Acknowledgements}

\smallskip

We thank S. Dimopoulos and J. Terning for interesting discussions.
NK thanks RESCEU, University of Tokyo, for kind hospitality during
the inception of this work. LS thanks the UC Davis HEFTI program
for hospitality in the course of this work. The research of NK is
supported in part by the DOE Grant DE-FG03-91ER40674. The research
of NK was also supported in part by a Research Innovation Award
from the Research Corporation. The work of LS is partially
supported by the U.S. National Science Foundation grant
PHY-0555304. The work of JY was partially supported by JSPS
Grant-in-Aid for Scientific Research Nos.~16340076 and 19340054.

~

%%%%%%%%%%%%%%%%%%%%%%%%%%%%%%%%%%%%%%%%%%%%%%%%%%%%%%%%%%%%%%%%%%%%%%%%

\end{document}